# Smooth and Low Latency Video Streaming for Autonomous Cars during Handover

Oussama El Marai[1] and Tarik Taleb[1,2,3]

[1] Aalto University, Espoo, Finland

[2] Centre for Wireless Communications (CWC), University of Oulu, Oulu, Finland

[3] Department of Computer and Information Security, Sejong University, Seoul, South Korea

*Abstract*—Self-driving vehicles are expected to bring many benefits among which enhancing traffic efficiency and reliability, and reducing fuel consumption which would have a great economical and environmental impact. The success of this technology heavily relies on the full situational awareness of its surrounding entities. This is achievable only when everything is networked, including vehicles, users and infrastructure, and exchange the sensed data among the nearby objects to increase their awareness. Nevertheless, human intervention is still needed in the loop anyway to deal with unseen situations or compensate for inaccurate or improper vehicle's decisions. For such cases, video feed, in addition to other data such as LIDAR, is considered essential to provide humans with the real picture of what is happening to eventually take the right decision. However, if the video is not delivered in a timely fashion, it becomes useless or likely produce catastrophic outcomes. Additionally, any disruption in the streamed video, for instance during handover operation while traversing inter-countries cross borders, is very annoying to the user and possibly cause damages as well. In this article, we start by describing two important use cases, namely Remote Driving and Platooning, where the timely delivery of video is of extreme importance [1]. Thereafter, we detail our implemented solution to accommodate the aforementioned use cases for self-driving vehicles. Through extensive experiments in local and LTE networks, we show that our solution ensures a very low end-to-end latency. Also, we show that our solution keeps the video outage as low as possible during handover operation.

## I. INTRODUCTION

The relatively fast connectivity offered by 4G networks in the last decade has essentially contributed to the explosion of data generated by both professionals and amateurs. Additionally, it opened the eyes for new services and applications (e.g. Augmented Reality, Self-driving Vehicles, and Remote Surgery) touching various fields, such as education, transportation and medical. These types of applications impose stringent requirements (less than 1ms for some use cases) and the violation of these requirements might lead to catastrophic results for verticals' businesses, and more importantly to human lives. The shift towards 5G is promising larger bandwidth, that could reach 20 Gbps which is 100 times faster than today's 4G LTE [2], and aims to incorporate Ultra-Reliable Low Latency Communications (URLLC), that theoretically drops to 1ms [3] to bolster the needs of these applications. To do so, the 5G architecture is adopting the novel concept of network slicing and dedicates a network slice to a specific use case or service (e.g. streaming).

The technology of autonomous cars is a highly active field that is receiving keen research attention from both industry and academia's researchers. Besides, governments and businesses are immensely investing billions of dollars in order to develop this technology and make it a reality [4]. Actually, different automobile industries and car manufacturers are scrambling to announce their first fully self-driving vehicles. The impetus behind is to reduce the accidents' risks caused by human failures, among which the erroneous estimation and the tiredness, offering more comfort while driving and providing a trustworthy and auguring ecosystem. Great economical revenues and eco-friendly outcomes are also expected.

Despite the fascinating features of autonomous cars and its prosperous future, the realization on the wild is confronting numerous challenges. For instance, every autonomous car should get high situational awareness of its surroundings, such as the weather, the moving objects and traffic signs and rules. This imposes the constant exchange of data flows in bi-directional ways (i.e. up- and down-link), which exerts high pressure on the underlying infrastructure in terms of consumed bandwidth and requires high processing capabilities to timely analyze and interpret the data.

Live video stream is one of the pivotal technologies that ongoing research is heavily relying on to accommodate many self-driving scenarios. Video data is known as bandwidth-consuming traffic due to the high size of the video frames even after compression [5]. Basically, a camera outputs a raw video stream of high resolution and size. This makes its conveyance over the network inefficient with nowadays networks. As a remediation, a video compression (e.g. using H.264 encoder) should be employed to reduce the size of the original stream size. Depending on many parameters, such as the used encoder, the target resolution, and bit-rate, this process introduces significant overhead to the video latency. Furthermore, the encoding is a time-consuming process and heavily uses CPU resources. Additionally, the protocol to be used (e.g. RTMP, RTSP, and HTTP) to convey the video stream over the network and the container format required by the player require further *Muxing* step. This would also add an overhead to the streaming latency. Adding to that, the time needed for the video data to flow to its destination and the inverse processes (i.e. Demuxing and Decoding processes) are done at the player side to visualize the stream in the receiver's screen. All these together, create the so-called *glass-to-glass*[1] latency that depends on many factors, such as the hardware specifications, the available bandwidth, and the technologies employed. This latency should be maintained as low as possible throughout the streaming session in order to meet the requirements of self-driving cars' use cases.

In this paper, we propose a solution for low latency live video streaming and study its applicability to both Remote Driving and Truck Platooning use cases for autonomous vehicles. Specifically, we propose the solution to minimize the video stream outage during the handover process between two different network operators (i.e., cross border) to ultimately provide smooth video playback.

The remainder of this paper follows the following structure. Previous work are summarized in the next section. Section III discusses the use cases to which this work applies. In Section IV, we describe our system architecture and detail its different components. Experimental results are provided in Section V. Finally, Section VI draws some concluding remarks.

## II. RELATED WORK

The advances in communication technologies have evolved enormously at a very fast pace over the last decade. These advances revolutionized our way of life in many domains, such as healthcare, automotive industry, and transportation. The emerging Intelligent Transportation Systems (ITS), and particularly the autonomous vehicles, is one of the many applications that have taken advantage of these advances to ultimately uphold safety, provide comfort for humans, while efficiently utilizing the roads and reducing fuel consumption. Vehicles' platooning concept is a promising solution to achieve the aforementioned goals and more. It has been extensively studied in many road traffic-related projects among which the California Partners for Advanced Transportation Technology (PATH) [6] in USA, Safe road trains for the environment (SARTRE) [7] and CHAUFFEUR [8] in Europe, and the KONVOI project [9] funded by the German government. In this section, we briefly review some of the previous studies devoted to the platooning concept.

In [10], the authors study the leader election problem in a platoon. Since the leading vehicle is responsible for most of the self-driving vehicles' functionalities and maneuvers (e.g., taking a turn, braking, speeding up and slowing down) in the whole platoon, it is very crucial to select the most qualified vehicle that will take this responsibility in such highly dynamic environment. To this end, the authors propose a consensus-based mechanism and adopt an incentive strategy for the leader election under different circumstances. In the same context, Jaehee et al. propose in [11] the use of the Raft algorithm for the election of a platoon leader. In the proposed strategy, all vehicles in the platoon can have one of the three states: *Followers*, *Leaders*, and *Candidates*. Initially, a vehicle gets *Follower* state once it joins the platoon and switches to Candidate when its random election timeout elapses, which triggers the election period. The better the performance of a vehicle, the faster the election timeout decreases. Then, Candidate vehicles send a request to the rest of the vehicles asking for a vote, after voting for themselves. The other vehicles vote for the one who sends the request earlier, and eventually the vehicle with the majority of votes wins.

Hao et al. tackle the problem of platoon head selection from the security standpoint in order to prevent choosing badly-behaving head vehicles [12]. To this aim, they have proposed the *REPLACE* solution consisting of a recommendation system for platoon head vehicles based on the vehicles' reputation collected from the user's feedback. In [13], a platoon management framework, based on Vehicular Ad-hoc Network (VANET) and CACC vehicles, has been proposed by Mani et al. In this framework, three main platoon operations, namely merge, split and lane change, are considered to accomplish the three scenarios of entry, leader leave, and follower leave, respectively. The communication between the vehicles within the platoon is achieved through V2V using Dedicated Short Range Communication (DSRC) having up to 1000 meters of transmission range. Simulation results, based on SUMO and ONMNET++, show the effectiveness of the proposed protocol in terms of traffic flow stability and throughput.

In [14], Uhlemann presents and discusses many recent research efforts conducted by major industries such as Audi, Volvo and Otto in the assisted driving field. The paper also describes the relevant projects and events, such as the Grand Cooperative Driving Challenge (GCDC), aiming to promote the concept of cooperative driving for vehicle automation. Uhlemann showcases the collaboration between large technological and communication companies (e.g. Cisco and Nokia) and car manufacturers (e.g. Hyundai) in different projects by leveraging recent technological advances such as IoT and UAVs to enhance the developments towards fully autonomous vehicles.

Mouelhi et al. propose in [15] a novel distributed autonomous control system for connected vehicle platoons using Distributed Object-Oriented Component-Based Design (DOOCBD). The proposed system permits controlling the speed of vehicles and avoiding collisions among vehicles in a platoon. Specifically, they leverage Vehicle-To-Everything (V2X) communication technologies to allow a new vehicle to join a vehicle platoon and to propagate a braking decision, taken by the leader vehicle in case of an obstacle for instance, to the rest of the vehicles in the platoon. A prototype of the proposed solution has been implemented and tested on a platoon of wheeled robots to demonstrate the feasibility and effectiveness of the proposed solution.

## III. LOW LATENCY USE CASES FOR AUTONOMOUS CARS

In this section, we describe two use cases in which the video feed is essential. Particularly, we describe the Remote Driving and Platooning use cases where the former enables a Remote Human Operator (RHO) to remotely take over the control of a vehicle, and the latter allows the drivers of the follower trucks in a platoon to get a better view of the roads. In both use cases, the latency of the live video stream should be maintained as low as possible (typically below 500ms) in order to provide a safe self-driving experience.

---

[1]It means from the camera's glass to the end-user screen's glass.

## A. Remote Driving Use Case

As its name infers, fully autonomous vehicles are supposed to operate safely and independently from human supervision. To do so, they should be able to first acquire all and every movement from their surrounding, second, and based on the received data, they should be able to take the right decision at the right time. While the first task is currently feasible, notably with the various highly accurate sensing devices, the second task remains elusive due to unforeseen situations. For such cases, the remote intervention of a human being becomes indispensable to either fully overtake the driving from the autonomous vehicle or to validate the vehicles' decision in case of doubts.

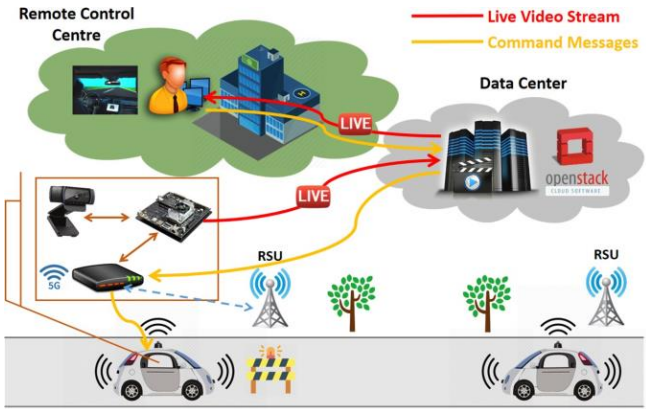

Fig. 1: Remote driving use case.

Remote Driving is an indispensable use case for self-driving cars. It allows the RHO at a Remote Control Center (RCC) to take over the control of a vehicle in some particular cases such as facing an unexpected obstacle during an already planned journey. To do so, a Single Onboard Computer (SOC) is deployed to the self-driving car and is connected to its Onboard Unit (OBU). Basically, OBU provides both the vehicle and the SOC with connectivity over 5G networks. A HD webcam is also connected to the SOC to stream the scene live. Upon detecting an unexpected event, the vehicle stops immediately and sends a remote driving request to the RCC. At the same time, it triggers a request to the SOC to start streaming the scene live along with the last 15 seconds video stream that precedes the request, so the RHO will get better understanding of the triggering event. It also sends other sensing data, such as the Lidar and the Real-Time Kinematic (RTK) positioning. Based on the video feeds and the different sensing data, the RHO sends commands to the vehicle (e.g. using a Joystick). Fig. 1 illustrates the overall architecture of the Remote Driving use case.

## B. See-What-I-See in Platooning Use Case

Platooning is one of the self-driving vehicles' use cases that is gaining ample traction in both industry and academia. Similarly to trains where many wagons are *physically* connected to each other, a platoon of self-driving vehicles consists of many vehicles *virtually* attached to each other and traveling at close proximity. The first vehicle in the chain, called leader vehicle, takes over the main driving functionalities and decisions, such as speeding up, braking and steering. The role of the rest of the vehicles in the same platoon, called follower vehicles, is to keep listening to the leader's decisions and react accordingly. The wide interest from the community about the platooning use case is justified by the numerous potential benefits it promises in terms of safety, environmental, and economical levels. Fig. 2 illustrates the platooning scenario during the handover operation.

The aim of the See-What-I-See (SWIS) use case is to provide the drivers at follower trucks with the road view of the leader truck. This would bolster a comfortable driving experience and prevent late reactions (e.g. braking) caused by road invisibility due to the closeness of the vehicles to each other. To this end, the leader vehicle keeps streaming live the road to a media server in the cloud. The follower vehicles in the platoon subscribe to the live video stream sent by the leader vehicle and start receiving the stream from the server. Crucially, the system should ensure a smooth playback especially during handover operation, for instance when crossing the borders between two countries, and should minimize the stream disruption period caused by any network outage.

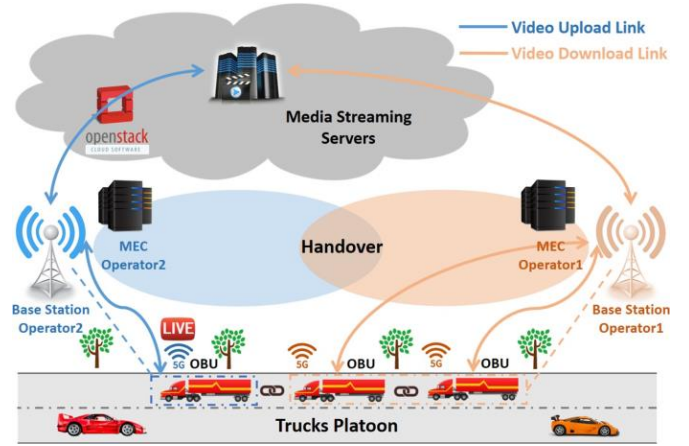

Fig. 2: See-What-I-See use case.

## IV. SYSTEM ARCHITECTURE

This section describes the proposed system architecture to enable low latency live video streaming for remote driving and platooning use cases in autonomous vehicles. Besides, the proposed architecture allows a smooth playback during Cross Borders Corridors (CBC) where vehicles change the Public Land Mobile Network (PLMN) when traversing from a country to another.

The proposed architecture is depicted in Fig. 3. As illustrated in the figure, there are three main streaming scenarios, namely Cloud-based, Fog-based and V2V-based streaming scenarios. For the two first scenarios, three entities (i.e., sender, server and receiver) are involved, while in the last one the sender plays the role of the server as well. In the Cloud-based scenario, the video stream is sent from the stream source to a central data center in the cloud, and the receiver(s) gets the stream through the server. This is more suitable for the remote driving use case where the RHO is located far away and there is no free RHO closer to the live video stream's sender. Fog-based streaming consists of leveraging the operator's Multi-access Edge Computing (MEC) resources mainly to reduce the streaming latency. This is achieved by deploying the streaming server closer to both the sender and receiver, hence minimizing the number of hops that the video packets could go through. This scenario is more adequate for the platooning use case since the platoon's members are geographically located in the same area. It could also be suitable for the remote driving use case when the RHO is in the proximity of the video stream's sender. To further reduce the latency, V2V-based streaming could be the best solution since it enables a direct connection between the sender and the receivers. In this case, the streaming server should run at the sender's side which will perform both encoding and delivering the stream. However, this solution applies only to the platooning use case.

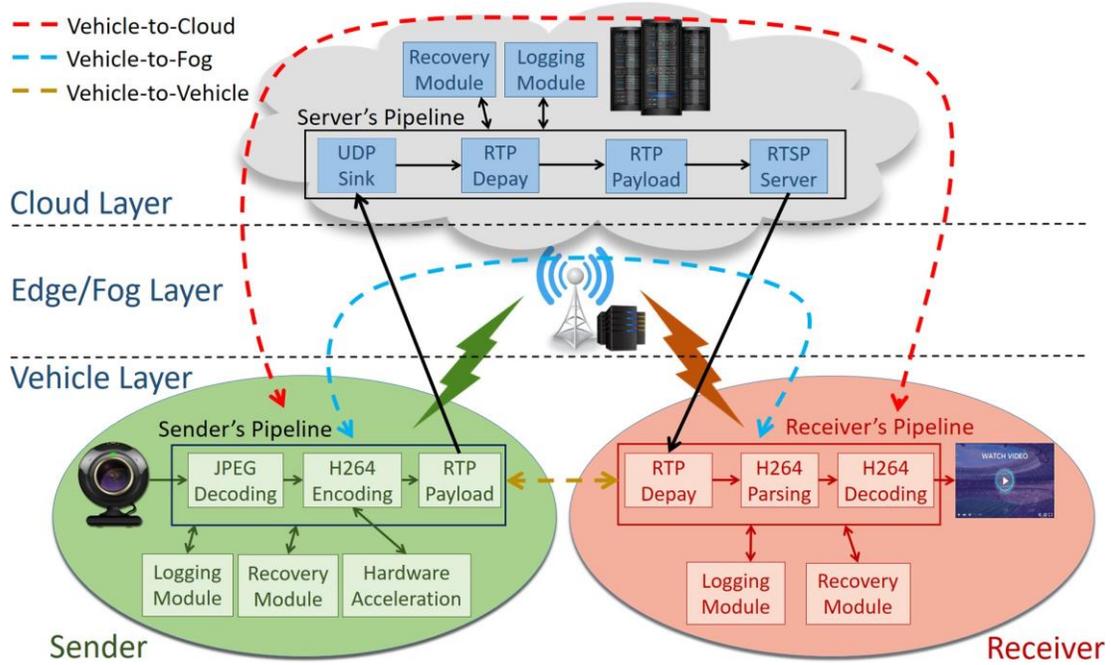

Fig. 3: Global system architecture.

As stated earlier, in order to enable platooning and remote driving use cases, three entities are required, namely the sender, receiver and the server. In the following, we describe each entity of the system as well as its corresponding pipeline elements.

### A. Stream Sender

The stream source is responsible for providing the live video stream captured from the vehicle's camera and deliv-ering it to the entity that has initiated the live video stream request. Specifically, in the platooning use case, one of the follower vehicles might be the initiator of the live video stream, while in the remote driving use case, the vehicle itself triggers the request when encountering an obstacle or unseen situation. Upon receiving the request, the SOC launches the live video streaming process and starts receiving the stream from the camera either in a raw or compressed format such as jpeg. To efficiently use the network's bandwidth and reduce the latency, it is necessary to encode the raw stream with highly efficient encoders such as Advanced Video Coding (AVC), also referred to as H.264, or High Efficiency Video Coding (HEVC), also known as H.265. In order to transmit the video stream over the network, a payload step should be added to encapsulate the encoded frames into RTP packets. These packets are sent over User Datagram Protocol (UDP) to a remote UDP sink. Generally, the encoding phase is known as a very time-consuming process. To speed up this process, GPU hardware encoding is leveraged to ultimately reduce the end-to-end (E2E) latency. To cope with failure cases, such as momentary network disruption, a recovery module is implemented. This recovery module keeps monitoring the whole pipeline and proceeds with recovery accordingly. A logging module allows measuring the sender's performance in terms of latency as well as resource consumption such as CPU, GPU, and RAM. The measurement of the latency is done at each media element in the pipeline.

### B. Streaming Server

The role of the streaming server is to deliver the stream to the receivers such as the follower vehicles or RHO. To do so, the server should run a media pipeline that receives the stream coming from the sender on a specific port and delivers it to the receivers over Real Time Streaming Protocol (RTSP). Similarly, logging and recovery modules are implemented to take measurements and recover from failures, respectively.

### C. Stream Receiver

The receiver of the stream could be either the follower vehicle or the RHO. It gets notified, through a REST API call, about the availability of the live video stream to launch the player's pipeline and initiate a streaming session with the server. Once the session is successfully established, it starts receiving the RTP packets and proceeds with depaying, parsing, decoding the compressed frames and finally displaying the raw stream on the user's screen. The recovery module is also present at the receiver side to ensure the stream's continuity and smoothness after handover operation.

## V. EXPERIMENTATION AND PERFORMANCE RESULTS

In this section, we describe the setup of our experimentation and discuss the achieved performance in two different setups. The first one consists of a Fog-based setup whereby we inter-connect three different machines via a router in a local wired network. In the second setup, both the sender and the receiver are connected through a 4G LTE network, and the server is located in an OpenStack cloud platform. The remainder of this section is organized as follows: we first discuss the technology choice, including the streaming protocol and the encoding/decoding tool. After that, we describe our evaluation setup, and then present the performance results achieved in each setup.

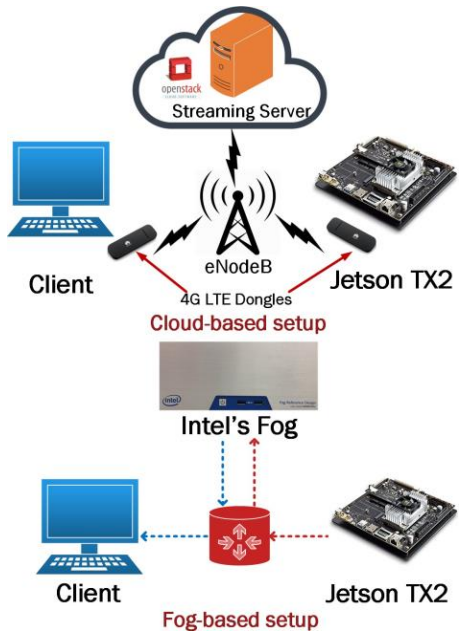

Fig. 4: Experiment setup.

### A. Technology Choice

To ensure a smooth playback of live video stream during handover (i.e., when traversing a cross borders corridor between two different countries), we first used Dynamic Streaming over HTTP (DASH) streaming technology [16]. DASH technology presents several advantages, among which it is a stateless protocol which permits overcoming the streaming outage during handover operations. We recall that after a successful handover operation, the client gets a new IP address from the target Public Land Mobile Network (PLMN). In this context, streaming over HTTP is very suitable to avoid getting disruptions since each video chunk is delivered in response to a separate HTTP GET request. Typically, the length of a chunk in DASH implementation is between 2 to 10 seconds. If we even reduce the chunk length to one second, this would introduce an extra delay that prevents keeping the latency as low as possible. Additionally, a 1-second chunk length will result in at least 1 second of latency which is not acceptable in both remote driving and platooning use cases.

Alternatively, we opted for Real Time Streaming Protocol (RTSP) [17] to meet the required latency for such use cases. Indeed, RTSP provides good results. However, it is a stateful protocol that results in session outage when the receiver joins the target PLMN due to IP address change. As a remediation, we deploy at each host a program written in C language that uses the GStreamer framework and keeps monitoring the host connectivity and accordingly recovers the streaming session after handover operation once the connection to the target PLMN is successfully established. To further reduce the latency, we use the hardware acceleration feature of NVIDIA GPU in the Jetson TX2 SOC for the encoding process. Basically, each program creates a pipeline of media elements where each of which is responsible for accomplishing a specific task such as acquiring, encoding, decoding, payloading, depaying and displaying the stream. The use of a SOC that has a GPU capability (e.g. Jetson TX2) is crucial to considerably lower the latency by leveraging the hardware capabilities, notably during the encoding phase, which is considered as the most resource-consuming process in the streaming pipeline. It is worth mentioning that the achieved results are obtained by employing hardware encoding at the sender only, and lower latency would certainly be achieved if the stream receiver has GPU capabilities to employ hardware decoding.

As to the encoding/decoding process, we have used GStreamer [18]. It is a powerful open-source, cross-platform multimedia framework with a programmable pipeline plugin architecture. Additionally, unlike the widely used FFmpeg, it supports many plugins for hardware encoding/decoding in NVIDIA GPUs deployed in SOCs.

### B. Platform Setup

The evaluation of the live video streaming service in case of the remote driving and platooning use cases during the handover process is done using a testbed composed of three hosts. The first host consists of a Jetson TX2 SOC having a GPU architecture with 256 NVIDIA CUDA cores, a Quad-Core ARM® Cortex® -A57 MPCore, and 8GB 128-bit LPDDR4 Memory. The Jetson TX2 runs Ubuntu 18.04. The second host is a virtual machine in an OpenStack cloud server

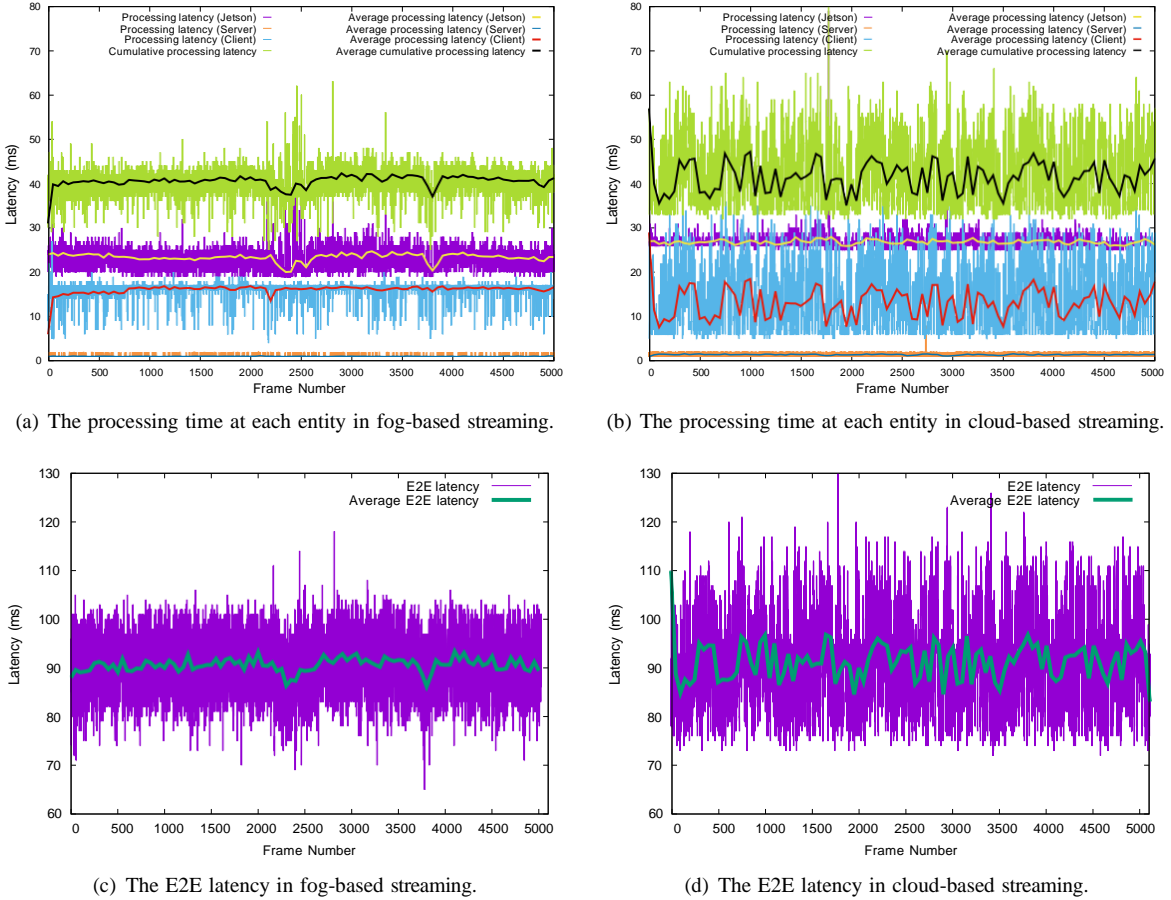

(a) The processing time at each entity in fog-based streaming.

(b) The processing time at each entity in cloud-based streaming.

(c) The E2E latency in fog-based streaming.

(d) The E2E latency in cloud-based streaming.

Fig. 5: The processing and E2E latency in both fog- and cloud-based streaming.

with 4 cores Intel Processor (Haswell, no TSX, IBRS) and 4GB of RAM, and is running Ubuntu Server 18.04. The last host represents the receiver machine where the player runs and displays the received stream. It consists of a desktop machine running Ubuntu Desktop 18.04. Its hardware configuration consists of an Intel(R) Xeon(R) CPU E31230 @ 3.20GHz, 8GB RAM. The experiments were conducted in both fog- and cloud-based setups. In the fog-based network setup, the three hosts are interconnected with each other through a physical router by an Ethernet cable, whilst in the cloud-based setup, both the sender and receiver are connected through 4G LTE dongles. Fig. 4 illustrates the experiment setup in both fog- and cloud-based live video streaming scenarios.

### C. Performance Results

In this section, we present the achieved results in terms of latency during the encoding and decoding processes in both fog- and cloud-based setups. We start by showing measurements regarding the computation time needed at each end. This is achieved by adding probes at each media element in the pipelines that are deployed at each host. Then, we present the measured E2E latency that includes the network propagation delay from the sender to the server as well as from the server to the receiver. We also present the results about the displayed and dropped frames over 11 different executions as well as the 5%-95% confidence interval of the latency. Finally, we show the consumed time to recover a stream session during a handover operation.

*1) Fog-based streaming latency:* In this scenario, we evaluate the live video streaming latency when the streaming server is in proximity to both the sender and receiver(s). The evaluation covers both the processing time and the measured E2E latency.

Fig. 5(a) shows the per-frame processing time needed at each streaming entity. The results show that the processing time needed for capturing, decoding, encoding with H264 and payloading with RTP format at the sender side using the Jetson TX2 is the part that takes more time compared to the server and receiver. Specifically, the encoding process is the phase that takes most of the time. At the server-side, we notice that the processing time for one frame does not take more than 2 ms since there is no encoding or decoding there. The last element in the chain is the receiver. It performs the depaying, decoding and displaying respectively. This takes less time than the sender's processing time. It should be noted here that the encoding at the sender is done using hardware acceleration, whilst the decoding at the receiver side is done using a software decoder. This figure also shows the

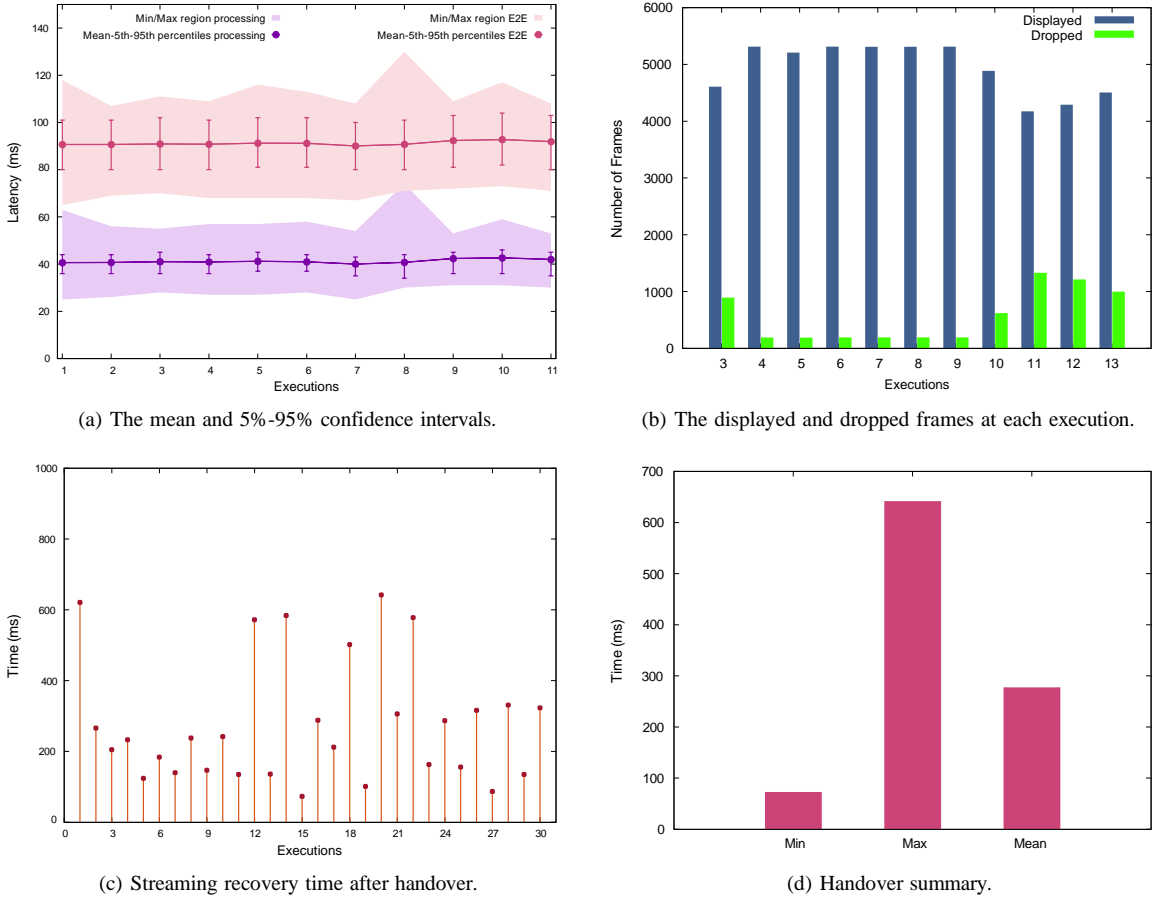

(a) The mean and 5%-95% confidence intervals.

(b) The displayed and dropped frames at each execution.

(c) Streaming recovery time after handover.

(d) Handover summary.

Fig. 6: Streaming recovery time during handover operation.

accumulative processing time of the three entities which falls within the interval [30, 50] ms most of the time. There are some exceptional outlier values generated from the encoding process. The average accumulative processing time is around 40ms most of the time.

The measured E2E latency, including the network, is depicted in Fig. 5(c). The results show that the network adds almost 50ms on average to the accumulative processing time. It should be noted that we used Network Protocol Time (NTP) to synchronize the three entities where the server is set to be the reference time for both the sender and receiver.

Fig. 6(a) depicts the 5%-95% confidence interval of the processing time and E2E latency in fog-based streaming experiment. These results are calculated after conducting 11 different executions. From this figure, we can vividly see that the overall mean latency is nearly the same at all executions, and the E2E latency is bounded and not oscillating which offers a stationary performance.

As we are using UDP as the transport protocol, some frames are dropped when they arrive late. Fig. 6(b) shows the number of displayed and dropped frames at each execution. From this figure, we observe that the dropping rate in most executions is very low, and is relatively high at few executions. This is mainly due to the network conditions during the experiment.

*2) Cloud-based streaming latency:* We have also conducted the same experiment but using 4G LTE dongles at both sender and receiver, as shown in Fig. 4. The measured results are plotted in Figs. 5(b) and 5(d). From Fig. 5(b), we observe almost no change in the processing time at the server side. However, the latency has increased for both the sender and receiver and it is noticeably unstable compared to the fog-based streaming experiment. We also notice from Fig. 5(d) that obviously the E2E latency increased by around 25ms in 4G LTE network compared to the fog-based streaming. Similarly, the E2E latency in cloud-based streaming is also bounded even though it is less stable than the fog-based experiment.

*3) Streaming session recovery after handover operation:* Resuming the stream automatically after a network outage due to handover operation is an important feature in our implemented solution. Additionally, the gap between the time the video freezes and it resumes playing back should be minimized as much as possible to offer smooth and seamless playback of the video stream, even after handover when the OBU gets another IP address which results in dropping the session between the receiver and the server. To this end, we have deployed a module that keeps monitoring the pipeline status and re-establishes a new session with the server once a handover operation is accomplished and a new IP address is obtained. In Fig. 6(c), we present the measured time to recover the streaming session after handover operation in 30 handover operations. The reported values include the time since losing the connection to the instant when a new session is re-established with the server. Fig. 6(d) summarizes the obtained results by presenting the minimum, mean and maximum values. The obtained results show that the average recovery time is lower than 300ms, and occasionally surpasses 600ms, which is usually unnoticeable by human eyes.

## VI. CONCLUSION

Autonomous vehicles are essential and promising technology for future smart cities. It brings many benefits in various domains among which environmental, economical and more importantly human safety by reducing accidents due to human errors on the roadways. Despite the ability of self-driving vehicles to handle many driving maneuvers such as automatic speed control, braking and overtaking, autonomous systems are prone to failures due to security attacks, unseen situations or improper vehicle decisions. Hence, human intervention should always be considered as an alternative solution. In this case, humans need data feed from the vehicle's sensors in order to be able to react when vehicles fail to take the right decision. To this end, live video streams are deemed to be the most suitable and interactive data feeds.

In this article, we presented two important use cases that require live video streams for human operators, namely *remote-driving* and *platooning*. In the first, the human operator is located in a remote control center and receives a take over driving request from the self-driving vehicle in case of obstacles or confused situation. Upon reception, the live video stream, as well as the last recorded 15 seconds video streams, are automatically streamed to the remote human operator. Contrary, in the platooning use case, the human is located in the follower vehicle of a platoon of vehicles. In this use case, the live video stream is provided mainly for comfort purposes, but also for taking quick action by the human driver when needed. In this context, we implemented and evaluated a live video streaming solution that ensures a low E2E latency. Additionally, the implemented solution ensures low streaming experience outage during handover operation.

In future work, we aim to study the streaming latency for various video qualities and resolutions including 4K with H.265 compression algorithm and its impact on the system performance such as the latency, bandwidth, the GPU, CPU and RAM usage. We will also investigate how to eliminate the delay introduced by the player at the receiver side in order to further lower the E2E latency.

## ACKNOWLEDGMENT

This work was supported in part by the Academy of Finland CSN project (Grant No. 311654) and the Academy of Finland Project 6Genesis Flagship (Grant No. 318927). It is also a part of the 5G-MOBIX project, that received funding from the European Union's Horizon 2020 research and innovation programme under grant agreement No 825496. Content reflects only the authors' view and the European Commission is not responsible for any use that may be made of the information it contains.

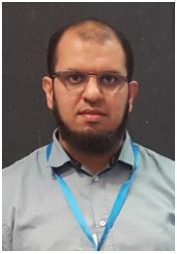

**Oussama EL Marai** received his Engineering degree in computer science in 2005 from University of Science and Technology Houari Boumediene (USTHB), Algiers, Algeria, Master degree from Ecole nationale Superieure d'Informatique (ESI) in 2009 and he is pursuing his doctoral degree at the School of Electrical Engineering, Aalto University, Finland. His research interests include adaptive video delivery, video QoE optimization, digital twin and image processing.

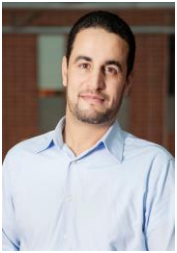

**Tarik Taleb** (Tarik.Taleb@aalto.fi) received the B.E. degree (with distinction) in information engineering in 2001, and the M.Sc. and Ph.D. degrees in information sciences from Tohoku University, Sendai, Japan, in 2003 and 2005, respectively. He is currently a professor with the School of Electrical Engineering, Aalto University, Espoo, Finland. He is the founder and the director of the MOSA!C Lab (http://www.mosaic-lab.org/).